\documentstyle[12pt]{article}
\def\be{\begin{eqnarray}}
\def\ee{\end{eqnarray}}
\def\ba{\begin{array}}
\def\ea{\end{array}}
\def\G{{\cal G}}

\def\A{{\cal A}}
\def\M{{\cal M}}

\def\S{{\cal S}}

\def\X{{\cal X}}
\def\Z{{\cal Z}}

\def\H{{\cal H}}

\def\Q{{\cal Q}}

\def\E{{\cal E}}
\def\F{{\cal F}}
\def\U{{\cal U}}
\def\V{{\cal V}}
\def\P{{\cal P}}
\begin{document}
\begin{center}
{\bf\LARGE {String theory extensions of Einstein-Maxwell fields:\\
         \vskip 3mm
	    the static case}}
\end{center}
\vskip 10mm

\begin{center}
{\large Alfredo Herrera-Aguilar\footnote{E-mail address:
{\tt herrera@itzel.ifm.umich.mx}},
Oleg V. Kechkin\footnote{E-mail address: {\tt kechkin@ginette.ifm.umich.mx},\,
{\tt kechkin@depni.npi.msu.su}}
\footnote{On leave from Institute of Nuclear Physics of
M.V. Lomonosov Moscow State University, Vorob'jovy Gory, 119899 Moscow, Russia}}
\end{center}

\vskip 5mm

\begin{center}
Instituto de F\'\i sica y Matem\'aticas\\
Universidad Michoacana de San Nicol\'as de Hidalgo\\
Apdo. Postal 2-82, Morelia, Mich., M\'exico
\end{center}

%

\vskip 5mm

\begin{abstract}
We present a new approach for generating solutions in both the
four-dimensional heterotic string theory with one vector field and the
five-dimensional bosonic string theory, starting from static
Einstein-Maxwell fields. Our approach allows one to construct classes
of solutions which are invariant with respect to the total
subgroup of three-dimensional charging symmetries of these string theories.
The new solution--generating procedure leads to the extremal
Israel-Wilson-Perjes subclass of string theory solutions in a special
case and provides its  natural continuous
extension to the realm of non-extremal solutions. We explicitly calculate all string
theory solutions related to three-dimensional gravity coupled to an effective dilaton
field which arises after an appropriate charging symmetry invariant reduction of the
static Einstein-Maxwell system.
\end{abstract}
\vskip 5mm
\begin{center}
\end{center}

\newpage

\renewcommand{\theequation}{\thesection.\arabic{equation}}
\section{Introduction}
\setcounter{equation}{0}
In string theories the study of the solution
spectrum for their effective field theory limits plays an important role
\cite{GSW}, \cite{Kir}, \cite{SenInt}, \cite{Youm}. This study includes both a
straightforward construction of
new solutions \cite{Kal}, \cite{BLS}, \cite{Sab}, \cite{Cve1}, \cite{Cve2}
\cite{ggk}
and various applications of symmetry based generation procedures
\cite{SenSol}, \cite{gk1}, \cite{ky}. The subject
of this paper is related to the former approach; in fact we propose a
new possibility of generating string theory solutions in an explicitly
symmetry invariant form
starting from the well studied system of  Einstein-Maxwell fields
(see \cite{EMT} and
the references therein). Namely, we indicate a surprising possibility
of extending the static solutions of the Einstein-Maxwell theory to the
realm of both the four-dimensional
heterotic string theory with one vector field \cite{Rog}, \cite{kyemda},
\cite{gkud},
\cite{gkmep} and the five-dimensional bosonic string theory \cite{kybs}.

These two theories can be considered together when studying
their extremal Israel-Wilson-Perjes type solutions \cite{ZF}; in this paper we
present the continuous generalization of the corresponding results to the field of
non-extremal solutions. We show that both the extremal and non-extremal solution
subspaces can be represented in the remarkable Einstein-Maxwell form. We
explore the close analogy between the heterotic (bosonic) string theory and the
Einstein-Maxwell system \cite{HKCS} as some clear leading principle in the study of the two
concrete string theories mentioned above. We also perform a consistent
charging
symmetry invariant reduction of the string theory inspired static Einstein-Maxwell
system to the effective three-dimensional Einstein-dilaton theory -- a
procedure which
relates, for example, the Schwarzshild black hole solution to the
Nordstrom-Reissner one \cite{EMT}.

The black hole physics seems to be the most promising field of
applications of the new
approach developed in this article. This statement is based, from one side,
on its hidden symmetry invariant property, which leads to the construction of symmetry
non-generalizeble asymptotically flat classes of solution, and,
from the other side,
on its close relation to the well studied field of classical black
hole solutions
in Einstein-Maxwell theory \cite{EMBH}. The answer to the natural question about the
extension of this new approach to the case of string theories with $d+n>2$
will be
given in a forthcoming publication \cite{m2}. The experience obtained in
the field of extremal solutions \cite{ZF} leads to the separation of
all these effective string theories (with arbitrary values of $d$
and $n$, excepting the case of $d=1$, $n=0$ which does not possess extremal
solutions and is equivalent to a double General Relativity system in the
stationary case) in two classes: the first class contains the two special
theories under consideration (with $d+n=2$), whereas the second one
encloses all the remaining theories (with $d+n>2$). In fact, the latter
systems can also be incorporated into the approach developed below
after some simple but important modifications of the presented
formalism.

\section{3D heterotic string theory: review of new formalism}
\setcounter{equation}{0}

In this paper we actively explore a new formalism developed in details in \cite{ZF}
for the general case of the $D$-dimensional low-energy heterotic string theory with $n$
Abelian gauge fields (the bosonic string theory corresponds to the special case of
$n=0$). We consider the toroidal compactification of this theory
to three spatial dimensions originally performed in
\cite{MahSch}--\cite{Sen3} and settled in a convenient form
in \cite{HKMEP}, \cite{HKCS} and \cite{ZF}. Let us briefly review some elements of
this new formalism which are necessary for the further analysis.

Thus, we start with
the action for the bosonic sector of the low-energy heterotic string theory
\cite{Kir}:
\be\label{2.1}
\S_D=\int d^DX |{\rm det}G_{MN}|^{\frac{1}{2}}\,e^{-\Phi}
\left ( R_D+\Phi_{,M}\Phi^{,M}-\frac{1}{12}H_{MNK}H^{MNK}-
\frac{1}{4}F^I_{MN}F^{I\,MN}\right ),
\ee
where $H_{MNK}=\partial_MB_{NK}-\frac{1}{2}A^I_MF^I_{NK}+\,{\rm cyclic}\,
\{M,N,K\}$ and $F^I_{MN}=\partial_MA^I_N-\partial_NA^I_M$. Here $X^M$ is the
M-th $(M=1,...,D)$ coordinate of the physical space-time of signature
$(-,+,...,+)$, $G_{MN}$ is the metric, whereas $\Phi$, $B_{MN}$ and $A^I_M$ \,
$(I=1,...,n)$ are the dilaton, Kalb-Ramond and Abelian gauge fields,
respectively. To determine the
result of the toroidal compactification to three dimensions, let us put $D=d+3$,
$X^M=(Y^m,\,x^{\mu})$ with $Y^M=X^m$\, $(m=1,...,d)$ and $x^{\mu}=X^{d+\mu}$
\, $(\mu=1,2,3)$ and introduce the $d\times d$ matrix $G_0={\rm diag}\,
(-1;1,...,1)$,
the $(d+1)\times (d+1)$ and $(d+1+n)\times (d+1+n)$ matrices $\Sigma$ and
$\Xi$ of the form ${\rm diag}\, (-1,-1;1,...,1)$, respectively,
and the $(d+1)\times (d+1+n)$ matrix field $\Z=\Z(x^{\lambda})$
together with the three-metric
$h_{\mu\nu}=h_{\mu\nu}(x^{\lambda})$. In \cite{ZF} it was shown that the
resulting theory after the
toroidal compactification of the first $d$ dimensions $Y^m$ can be expressed in terms
of the pair $(\Z,\,h_{\mu\nu})$; its effective dynamics is given by the action
\be\label{2.2}
\S_3=\int d^3x h^{\frac{1}{2}}\left ( -R_3+L_3\right ),
\ee
where $R_3=R_3(h_{\mu\nu)}$ is the curvature scalar for the three-dimensional line
element $ds_3^2=h_{\mu\nu}dx^{\mu}dx^{\nu}$ and
\be\label{2.3}
L_3={\rm Tr}\,\left [\nabla\Z\left (\Xi-\Z^T\Sigma\Z\right )^{-1}
\nabla\Z^T\left (\Sigma-\Z\Xi\Z^T\right )^{-1}\right ].
\ee
It is important to note that in the present notations $Y^1$ corresponds
to time, thus, we
consider stationary equations of motion and $x^{\mu}$ are the coordinates on the
Riemannian three-dimensional space.
To translate this ($\sigma$-model) description into the language of the
field components of the heterotic string theory, let us introduce three doublets of $(\Z,\,h_{\mu\nu})$-related
potentials $(\M_{\alpha},\vec\Omega_{\alpha})$ \, $(\alpha=1,2,3)$ according to the
relations
\be\label{2.4}
&&\M_1=\H^{-1},\quad \nabla\times\vec\Omega_1=\vec J,
\nonumber\\
&&\M_2=\H^{-1}\Z,\quad \nabla\times\vec\Omega_2=\H^{-1}\nabla\Z-\vec J\Z,
\nonumber\\
&&\M_3=\Z^T\H^{-1}\Z,\quad \nabla\times\vec\Omega_3=\nabla\Z^T\H^{-1}\Z-\Z^T\H^{-1}
\nabla\Z+\Z^T\vec J\Z,
\ee
where $\H=\Sigma-\Z\Xi\Z^T$ and $\vec J=\H^{-1}\left (\Z\Xi\nabla\Z^T-\nabla\Z\Xi\Z^T\right )
\H^{-1}$. In Eq. (\ref{2.4}) the scalars $\M_{\alpha}$ are off-shell quantities, whereas
the vectors $\vec\Omega_{\alpha}$ are defined on-shell. The scalar and vector potentials
forming any doublet have the same matrix dimensionalities; let us represent them in
the following block form
\be \label{2.5}
\left(
\ba{cc}
1\times 1&1\times d\cr
d\times 1&d\times d
\ea
\right),
\quad
\left(
\ba{ccc}
1\times 1&1\times d&1\times n\cr
d\times 1&d\times d&d\times n
\ea
\right),
\quad
\left(
\ba{ccc}
1\times 1&1\times d&1\times n\cr
d\times 1&d\times d&d\times n\cr
n\times 1&n\times d&n\times n
\ea
\right)
\ee
for $a=1,2,3$ respectively, where, for example, the
`$13$' block components of the potentials $\M_2$ and $\vec\Omega_2$ are
$1\times n$ matrices. Afterwards, let us define the following set of
scalar and vector quantities:
\be \label{2.6}
&&S_0=-\M_{1,11}+2\M_{2,11}-\M_{3,11},
\nonumber\\
&&S_1=G_0\M_{1,22}G_0+G_0\M_{2,22}+(\M_{2,22})^TG_0+\M_{3,22}+S_0^{-1}
\left [-\M_{1,12}G_0-\M_{2,12}+
\right.
\nonumber\\
&&\left. +
(\M_{2,21})^TG_0+\M_{3,12}\right ]^T
\left [-\M_{1,12}G_0-\M_{2,12}+(\M_{2,21})^TG_0+\M_{3,12}\right ],
\nonumber\\
&&S_2=G_0\M_{1,22}-G_0\M_{2,22}G_0+(\M_{2,22})^T-\M_{3,22}G_0-1+S_0^{-1}
\left [-\M_{1,12}G_0-\M_{2,12}+
\right.
\nonumber\\
&&\left. +
(\M_{2,21})^TG_0+\M_{3,12}\right ]^T
\left [\M_{1,12}-\M_{2,12}G_0+(\M_{2,21})^T-\M_{3,12}G_0\right ],
\nonumber\\
&&S_3=\sqrt 2\left [
G_0\M_{2,23}+\M_{3,23}+S_0^{-1}
\left (-\M_{1,12}G_0-\M_{2,12}+(\M_{2,21})^TG_0+\M_{3,12}\right )^T\times
\right.
\nonumber\\
\ &&\left. \times
\left (\M_{2,13}+\M_{3,13}\right )
\right ];
\nonumber\\
&&\vec V_1=\left [ -\vec\Omega_{1,12}G_0+\vec\Omega_{2,12}+(\vec\Omega_{2,21})^TG_0
+\vec\Omega_{3,12}\right ]^T,
\nonumber\\
&&\vec V_2=\left [ -\vec\Omega_{1,12}G_0-\vec\Omega_{2,12}G_0+(\vec\Omega_{2,21})^T
-\vec\Omega_{3,12}G_0\right ]^T,
\nonumber\\
&&\vec V_3=\sqrt 2\left ( \vec\Omega_{2,13}+\vec\Omega_{3,13}\right )^T.
\ee
In terms of them the heterotic string theory fields read \cite{ZF}:
\be\label{2.7}
&&ds^2_D=ds^2_{d+3}=(dY+V_{1\mu}dx^{\mu})^TS_1^{-1}(dY+V_{1\nu}dx^{\nu})+S_0ds_3^2,
\nonumber\\
&&e^{\Phi}=\left | S_0\,{\rm det}\,S_1\right |^{\frac{1}{2}},
\nonumber\\
&&B_{mk}=\frac{1}{2}\left ( S_1^{-1}S_2-S_2^TS_1^{-1}\right )_{mk},
\nonumber\\
&&B_{m\,d+\nu}=\left \{ V_{2\nu}+\frac{1}{2}\left ( S_1^{-1}S_2-
S_2^TS_1^{-1}\right )V_{1\nu}-S_1^{-1}S_3V_{3\nu}\right \}_{m},
\nonumber\\
&&B_{d+\mu\,d+\nu}=\frac{1}{2}\left [ V_{1\mu}^T\left ( S_1^{-1}S_2-
S_2^TS_1^{-1}\right )V_{1\nu}+V_{1\mu}^TV_{2\nu}-V_{1\nu}^TV_{2\mu}\right ],
\nonumber\\
&&A^I_m=\left ( S_1^{-1}S_3\right )_{mI},
\nonumber\\
&&A^I_{d+\mu}=\left ( -V_{3\mu}+S^T_3S^{-1}_1V_{1\mu}\right )_I.
\ee

At the end of this section we would like to make two remarks. The first one concerns
the procedure of constructing the solutions and their final representation;
namely, it is convenient to perform the basic construction of the
solution in terms of $\Z$ and
$h_{\mu\nu}$ and leave the result in the (\ref{2.7}) form giving explicit
expressions for the quantities $S_{\alpha},\,\vec V_{\alpha}$ and
$S_0, S_1^{-1}, {\rm det}\,S_1$. The
second one is related to the hidden symmetries of the compactified theory
(\ref{2.2})-(\ref{2.3}). It is easy to see that the transformation
\be\label{2.8}
\Z\rightarrow
C_1\Z C_2
\ee
is a symmetry if $C_1^T\Sigma C_1=\Sigma$ and $C_2^T\Xi C_2=\Xi$. In
\cite{HKCS} and \cite{ZF} it was shown that this symmetry coincides with the total
group of the three-dimensional charging symmetries of the theory. This group does not
affect the trivial spatial asymptotics of the fields and forms the base
for the
three-dimensional generation technique of asymptotically flat solutions of the
theory. The $\Z$-representation, being the matrix potential representation
with the
lowest possible matrix dimensionality (the $\sigma$-model (\ref{2.2})-(\ref{2.3})
is in fact a symmetric space model parameterizing the coset
$O(d+1,d+1+n)/O(d+1)\times O(d+1+n)$, see \cite{Sen3} and \cite{ZF})
provides a
linear realization of the charging symmetry transformations. These facts
allow one to construct asymptotically flat classes of solutions in an
explicitly
charging symmetry invariant form. In the next sections we give concrete
illustrations to the remarks formulated above.

\section{String theories from static Einstein-Maxwell system}
\setcounter{equation}{0}

As it was pointed out in the Introduction, the close analogy between the heterotic (bosonic)
and the Einstein-Maxwell theories will be used as certain underlying
principle in the
study of the two concrete string theories under consideration. This analogy, or
correspondence, was originally indicated in \cite{HKCS}; its explicit off- and
on-shell status was established in \cite{ZF}, whereas some natural applications were
given in \cite{OKPL1}, \cite{SU} and \cite{HSTI}. To formulate the correspondence in
appropriate terms, let us represent the Einstein-Maxwell theory in a form very
similar to the heterotic (bosonic) string theory one (\ref{2.2})-(\ref{2.3}). Namely,
let $E$ and $F$ be the conventional Ernst potentials of the stationary
Einstein-Maxwell theory \cite{Ernst}. Let us express the corresponding
three-dimensional Lagrangian
\be\label{3.1}
L_3=L_{EM}=\frac{1}{2f^2}\left | \nabla E-\bar F\nabla F\right |^2-
\frac{1}{f}\left | \nabla F\right |^2,
\ee
where $f=\frac{1}{2}(E+\bar E-|F|^2)$, in terms of the $1\times 2$ complex potential
$z$ with
\be\label{3.2}
z=(z_1\,\,z_2)
\ee
and (compare with \cite{Maz})
\be\label{3.3}
z_1=\frac{1-E}{1+E},\quad z_2=\frac{\sqrt 2F}{1+E}.
\ee
The result reads:
\be\label{3.4}
L_{EM}=2\frac{\nabla z(\sigma_3-z^+z)^{-1}\nabla z^+}
{1-z\sigma_3z^+},
\ee
where $\sigma_3={\rm diag}\,(1\,-1)$. Now, comparing Eqs. (\ref{2.3}) and
(\ref{3.4}), it is easy to see that the map
\be\label{3.5}
\Z\longleftrightarrow z,\quad \Xi\longleftrightarrow\sigma_3,\quad
\Sigma\longleftrightarrow 1,
\ee
together with the operation interchange $^T\leftrightarrow ^+$, relates the
heterotic (bosonic) and Einstein-Maxwell theories up to the factor
`2'. However,
this factor becomes necessary when considering the explicit on-shell realization of the
correspondence (\ref{3.5}) for the string theories with $d+n>2$.
Namely,
it is possible to identify the Einstein-Maxwell theory with some special truncation of
the string theory with
$d=1$ and $n=2$ \cite{OKPL1}, and after that, to extend this result to the
case of an
arbitrary theory with $d+n>2$ in a charging symmetry
invariant form
using the projectional formalism developed in \cite{HSTI}.
The above mentioned special truncation is in fact a consistent ansatz
with the $2\times 4$ potential $\Z_*$ parameterized as
\be\label{3.6}
\Z_*=(\Z_{*1}\,\,\Z_{*2})
\ee
where the block components $\Z_{*a}$ ($a=1,2$) are taken in the form
\be\label{3.7}
\Z_{*a}=
\left(
\ba{cc}
z_a^{'}&z_a^{''}\cr
-z_a^{''}&z_a^{'}
\ea
\right),
\ee
and the complex functions $z_a=z_a^{'}+iz_a^{''}$ correspond to Eq. (\ref{3.2}). The
statement is that in this special case, the heterotic string and the Einstein-Maxwell
theories coincide on-shell; in particular the Lagrangians
(\ref{2.3}) and (\ref{3.4}) are equal if Eqs. (\ref{3.6}) and (\ref{3.7}) take place.
Finally, the extension of stationary Einstein-Maxwell theory to the case of heterotic
string theory with $d+n>2$ is given by the map
\be\label{3.7'}
\Z_*\rightarrow \Z=L\Z_*R^T,
\ee
where $L^T\Sigma L=\Sigma_*,\,\,R^T\Xi R=\Xi_*$ and the matrices $\Sigma_*,\Xi_*$ and
$\Sigma,\Xi$ correspond to the theories with $\Z_*$ and $\Z$,
respectively.
It is easy to see that for the two exceptional
string theories with $d+n=2$ this
exact on-shell realization  (\ref{3.6})-(\ref{3.7'}) of
the correspondence (\ref{3.5}) becomes impossible, thus, in this case
the relations
(\ref{3.5}) possess a formal character. However, as it is shown
below, a surprising on-shell
correspondence between the string theories with $d+n=2$ and the Einstein-Maxwell
system nevertheless exists if we appropriately truncate both the string and
Einstein-Maxwell theories.

The necessary truncation naturally arises in the framework of the continuous
generalization of the general extremal Israel-Wilson-Perjes class of solutions
established for the string theories with arbitrary values of the parameters $d$
and $n$ in \cite{ZF}. This class can be represented in terms of the ansatz
\be\label{3.8}
\Z=\Lambda\Q,
\ee
where $\Lambda$ is a $(d+1)\times 1$ matrix function and $\Q$ stands
for a
$1\times 3$ constant matrix parameter for the theories with $d+n=2$. The class of
extremal Israel-Wilson-Perjes solutions arises when the (numerical) parameter
$\kappa$, defined by the relation
\be\label{3.9}
\kappa=\Q\Xi\Q^T,
\ee
vanishes, the dynamical quantity $\Lambda$ is harmonic and the three-metric
$h_{\mu\nu}$ is flat. In the Einstein-Maxwell theory the situation is
extremely
similar to this one according to the correspondence rule (\ref{3.5}):
the ansatz $z=\lambda q$,
where $\lambda$ is a complex function and $q$ is a $1\times 2$
constant complex
parameter, gives the conventional Israel-Wilson-Perjes class of solutions if the
parameter $\kappa=q\sigma_3q^+$ iz zero, $\lambda$ is harmonic and the three-metric
is again flat. The consideration of this ansatz in the case of an
arbitrary value of
$\kappa$ naturally leads to a continuous generalization of the extremal class of
Israel-Wilson-Perjes solutions to non-extremal classes. In the
framework of this
continuous extension, both extremal and non-extremal
Kerr-Newman-NUT solutions with the electric and magnetic charges belong to the same
family of solutions.
The main idea of this article is to perform the corresponding continuous
generalization of extremal classes of solutions for string theories with $d+n=2$
to non-extremal ones by considering the ansatz (\ref{3.8}) with a
non-zero value of the
parameter $\kappa$ (\ref{3.9}).

This consideration must be performed on-shell.
The straightforward substitution of the ansatz (\ref{3.8}) into the
equations
of motion derived from Eqs. (\ref{2.2}) and (\ref{2.3}), leads
to the following system of equations:
\be\label{3.11}
&&\nabla^2\Lambda+2\kappa\nabla\Lambda\Lambda^T\left (\Sigma-\kappa\Lambda
\Lambda^T\right )^{-1}\nabla\Lambda=0,
\nonumber\\
&&R_{3\,\mu\nu}=\kappa{\rm Tr}\,\left [\Lambda_{,(\mu}
\left ( 1-\kappa\Lambda^T\Sigma\Lambda\right )^{-1}\Lambda^T_{,\nu)}
\left (\Sigma-\kappa\Lambda\Lambda^T\right )^{-1}\right ].
\ee
It is clear that in the case $\kappa=0$ we come back to the extremal case studied
in \cite{ZF}, whereas for $\kappa\neq 0$ we have the above announced continuous
extension of the formalism to the non-extremal case. Below, in this section, we study
the situation
in which $\kappa\neq 0$; here the equations (\ref{3.11}) correspond to the effective
Lagrangian
\be\label{3.12}
L_{eff}=\kappa{\rm Tr}\,\left [
\nabla\Lambda
\left ( 1-\kappa\Lambda^T\Sigma\Lambda\right )^{-1}\nabla\Lambda^T
\left (\Sigma-\kappa\Lambda\Lambda^T\right )^{-1}\right ].
\ee
This Lagrangian has a quasi string theory form.
To clarify this statement, let us introduce the new dynamical variable
\be\label{3.13}
\zeta=\left (-\kappa\right )^{\frac{1}{2}}\Lambda^T
\ee
and consider the special case $\kappa<0$.
Let us also unify the two constrained string theories, i.e. the $d=n=1$ and
$d=2,\,\,n=0$ models restricted by Eq. (\ref{3.8}), into a single construction,
which in fact will correspond to the
former system. For this system $\Sigma=\Xi={\rm diag}\,(-1,-1,1)$ and
$\zeta=(\zeta_1, \zeta_2, \zeta_3)$. An evident statement is that the latter system
arises in the special case when $\zeta_3=0$. Thus, it is possible to consider the
$d=n=1$ heterotic string theory ansatz as a (consistent) subsystem
of the $d=2,\,\,n=0$ bosonic string theory one. Finally, both systems are
simultaneously described by a single effective theory
%
which, however, is not exactly of the form (\ref{2.3}) in view
of the fact that the string matrix potential $\Z$
must contain at least two rows. Interestingly, both these problems can be solved
by the formal introduction of the following $2\times 3$ potential
\be\label{3.15}
\Z=\left(
\ba{c}
\zeta \cr
0
\ea
\right)
\ee
together with the corresponding matrices $\Sigma$ and $\Xi$. In these terms the
effective theory
takes the form of Eq. (\ref{2.3}) and can be interpreted in the
framework of the $d=n=1$ heterotic string
theory
restricted by the relation (\ref{3.15}) which is a
consistent ansatz for the $d=n=1$ theory.

Now it is possible to
apply the Ernst matrix potential approach \cite{HKCS} to obtain an
appropriate
interpretation of this effective theory in terms of the well known
classical systems of gravity. This approach is based on the use of a pair
of Ernst matrix potentials $\X$ and $\A$ defined by the relations
\be\label{3.17}
\X=2\left ( \Z_1+\Sigma\right )^{-1}-\Sigma, \quad
\A=\sqrt 2\left ( \Z_1+\Sigma\right )^{-1}\Z_2,
\ee
where $\Z=(\Z_1\,\,\Z_2)$, compare to Eqs. (\ref{3.6})-(\ref{3.7}). In
terms of
these potentials the Lagrangian (\ref{2.3}) takes the form
\be\label{3.18}
L_3={\rm Tr}\,\left [
\frac{1}{4}\left(\nabla\X-\nabla\A\A^T
\right)\G^{-1}
\left(\nabla \X^T-\A\nabla \A^T\right)
\G^{-1}+\frac{1}{2}\G^{-1}\nabla\A\nabla\A^T
\right ],
\ee
where $\G=\frac{1}{2}(\X+\X^T-\A\A^T)$. By substituting (\ref{3.15})
into (\ref{3.17}) and then performing the calculations, we arrive to
the effective Lagrangian
\be\label{3.19}
L_{eff}=
\frac{1}{4\F^2}\nabla\F^2
+\frac{1}{\F}\left ( \nabla\V^2-\nabla\U^2\right ),
\ee
where the scalar potentials $\F, \V$ and $\U$ are
\be\label{3.20}
\F=\frac{1-\zeta_1^2-\zeta_2^2+\zeta_3^2}{(1-\zeta_1)^2},\quad
\V=\frac{\zeta_2}{1-\zeta_1},\quad
\U=\frac{\zeta_3}{1-\zeta_1}.
\ee

A classical interpretation of the effective theory is now clear:
the quantity $|\F|$ can be considered as the $-G_{tt}$ component
of the static Einstein-Maxwell theory, whereas $\sqrt 2\V$ and
$\sqrt 2\U$, as the electric and/or magnetic potentials of that theory.
Thus, in the framework of this interpretation one immediately obtains
\be\label{3.21}
L_{eff}=
\frac{1}{2}L_{EM},
\ee
where $L_{EM}$ is the corresponding Einstein-Maxwell Lagrangian in the
static case. Note that for $\F>0$ ($\F<0$) the potential $\V$ ($\U$)
must be imaginary and $\U$ ($\V$) must be real in order to perform such
identification. One also must remember that for the $d=n=1$ theory
$\U=0$, whereas for the $d=2$, $n=0$ model this dynamical variable is
not restricted. Let us now considrer the case $\U=0$ for both string
theories. If $\F<0$ we have the conventional interpretation of the
effective theory in terms of the static electric (magnetic)
Einstein-Maxwell system; if $\F>0$ the potential $\V$ becomes
essentially imaginary. Actually, the conventional static
Einstein-Maxwell theory with either electric or magnetic potential
corresponds to an indefinite $\sigma$--model, whereas the case under
consideration ($\F>0$), to a positive definite one.

An interpretation of the effective theory with $\U=0$ and $\F>0$ in
terms of well known gravity models
with strictly real fields takes place when
one identifies $\F^{\frac{1}{2}}$ and $\V$ with the $-G_{tt}$ and
the rotational metric component, respectively. In turns out that in
this case one can establish the following relationship
\be\label{3.22}
L_{eff}=
2L_{GR},
\ee
where $L_{GR}$ is the conventional Lagrangian of stationary
General Relativity. Note that the two special cases considered above
(with negative and positive values of $\F$) are related by the map
\be\label{3.22}
-\F\longleftrightarrow \F^{\frac{1}{2}}, \qquad
\V\longleftrightarrow i\V,
\ee
which is nothing else than the classical Bonnor transformation
\cite{B}. A last remark related to the factors `$\frac{1}{2}$' and `2'
is in order: when considering the axisymmetric theory these factors lead
to the maps $\gamma\longleftrightarrow \frac{1}{2}\gamma$ and
$\gamma\longleftrightarrow 2\gamma$, respectively, where
$\gamma=\gamma(\rho,z)$ enters the line element in the
Lewis--Papapetrou form
\be
ds_3^2=e^{\gamma}\left(d\rho^2+dz^2\right)+\rho^2d\varphi ^2;
\ee
thus, such maps do not affect our identifications.

At the end of this section let us establish all the hidden symmetries
of the effective theory (3.17). In order to do this, let us introduce the
$2\times 2$ matrix
\be
M=
\left(
\ba{cc}
f^{-1}&f^{-1}\chi_r\cr
f^{-1}\chi_l&f+f^{-1}\chi_l\chi_r
\ea
\right),
\ee
where $f=|\F|^{\frac{1}{2}}$, $\chi_r=\V+\U$ and
$\chi_l={\rm sign}(\F)(\V-\U)$. It is easy to see that this matrix
parametrizes the group $SL(2,{\bf R})$; a less obvious fact is that
\be
L_{eff}=\frac{1}{2}Tr\left(\nabla MM^{-1}\right)^2,
\ee
i.e., our complete ($\U\neq 0$) effective theory (3.17) coincides with the
$SL(2,{\bf R})$ principal chiral model coupled to gravity. From this
fact it immediately follows that the group of symmetry transformations
acts as
\be
M\longrightarrow C_l^TMC_r,
\label{3.24'}
\ee
where the unimodular constant matrices $C_{l,r}$ can be obtained from
$M$ by making use of the substitutions $f\longrightarrow s_{l,r}$,
$\chi_l\longrightarrow \beta_{l,r}$ and $\chi_r\longrightarrow \alpha_{l,r}$.
It is worth noticing that the transformations $C_l$ and $C_r$ are
independent each other, thus, there is no relation between the $-l$
and $-r$  labeled constant parameters $s_{l,r}$, $\alpha_{l,r}$ and
$\beta_{l,r}$. For example, one can fix $C_l=1$ and then write down the
transformations that correspond to the parameters $s_{r}$,
$\alpha_{r}$ and $\beta_{r}$; the result reads:
\be
f\longrightarrow s_r f, \qquad \chi_r\longrightarrow s_r^2 \chi_r, \qquad
\chi_l\longrightarrow \chi_l;
\ee
\be
f\longrightarrow f, \qquad
\chi_r\longrightarrow \chi_r+\alpha_r, \qquad
\chi_l\longrightarrow \chi_l;
\ee
\be
f\longrightarrow \frac{f}{1+\beta_r\chi_r}, \qquad
\chi_r\longrightarrow \frac{\chi_r}{1+\beta_r\chi_r}, \qquad
\chi_l\longrightarrow \frac{\chi_l+\beta_r\left(f^2+\chi_l\chi_r\right)}
{1+\beta_r\chi_r}.
\ee
Here the transformation (3.26) plays the role of scaling,
whereas (3.27) and (3.28) are shift and Ehlers--type maps, respectively.
Of course the matrix $C_r$ generates the transformations which can be
obtained from (3.26)--(3.28) using the interchange
$r\longleftrightarrow l$ in these relations.

Let us now discuss the special case corresponding to the subgroup of
charging symmetry transformatoins. This subgroup preserves the trivial
solution of the theory under consideration. Such a solution corresponds
to the matrix $M=M_0=1$ since for this special value one obtains
$\Z_0=0$ (see Eqs. (3.13)--(3.14),(3.18) and (3.23)).
Thus, for the subgroup of charging symmetries one obtains
$C_l=C_r^{T\,\,-1}$ from Eq. (\ref{3.24'}), or, in terms of the
corresponding parameters
\be
s_l=
\frac{s_r}{s_r^2+\alpha_r\beta_r}, \quad
\alpha_l=-\frac{\beta_r}{s_r^2+\alpha_r\beta_r}, \quad
\beta_l=-\frac{\alpha_r}{s_r^2+\alpha_r\beta_r}.
\ee

Some remarks on the $d=n=1$ string theory are in order.
It is clear that the effective theory with $\U\neq 0$ corresponds
to the string theory with $d=2$ and $n=0$. The theory with $d=n=1$
implies $\U\equiv 0$, which in turn leads to $\chi_l=\chi_r\equiv\chi$, and
thus, $M=M^T$ when $\F>0$. Thus, we deal with
the symmetric space model $SL(2,{\bf R})/SO(2)$ coupled to gravity
which is, in fact, equivalent to the stationary General Relativity
theory up to a factor `2' as it was explained above. The transformation
rule (\ref{3.24'}) must preserve the symmetric property of $M$, this
implies that
$C_l=C_r\equiv C\in SL(2,{\bf R})$. For these symmetric space the parameters
$s_l=s_r\equiv s$, $\alpha_l=\alpha_r\equiv\alpha$ and
$\beta_l=\beta_r\equiv\beta$
define the conventional scaling, shift and Ehlers maps. On the other
hand, the one--parameter charging symmetry subgroup is given by the
transformation
\be
z\longrightarrow e^{i\epsilon}z,
\ee
where $z=\frac{1-\E}{1+\E}$, $\E=f+i\chi$, $\epsilon=2\arctan\beta$
(the remaining parameters are related to $\beta$ as follows
$\alpha=-\beta$ and $s=\sqrt{1+\beta^2}$).

We hope to use the principle chiral model (3.23)--(3.24) for the generation of
new solutions in a forthcoming article. In the next section we perform
a further reduction of the effective theory (3.17) to the $\sigma$--model
with a single dynamical variable in a charging symmetry invariant form.
Our goal is to obtain an effective dilaton gravity system with
arbitrary coupling which allows one to consider the general field
configurations of Nordstrom--Reissner type in both the $d=2$, $n=0$
and the $d=n=1$ string theories.

\section{Explicit solutions via 3D dilaton gravity}
\setcounter{equation}{0}

In order to perform this truncation in the most general form, let us come back
to Eq.
(\ref{3.11}) and consider the new consistent ansatz
\be\label{4.1}
\Lambda=\Psi\P^T,
\ee
where $\Psi$ is a dynamical function and $\P$ is a $1\times ( d+1)$ constant row.
The resulting system of equations of motion reads:
\be\label{4.2}
&&\nabla^2\phi=0,
\nonumber\\
&&R_{3\,\,\mu\nu}=\sigma\phi_{,\mu}\phi_{,\nu},
\ee
where we have set $\sigma=\tau\kappa$,
\be\label{4.2'}
\tau=\P\Sigma\P^T
\ee
and
\be\label{4.3}
\Psi=\frac{\tanh{(\sigma^{\frac{1}{2}}\phi)}}{\sigma^{\frac{1}{2}}}.
\ee
The parameter $\sigma$ plays the role of the dilaton-gravity coupling; we identify
the field $\phi$ with the effective three-dimensional dilaton interacting with the
three-metric field $h_{\mu\nu}$ according to the equations (\ref{4.2}).
Note that, in general, $\sigma$ has arbitrary sign and for any sign of
$\sigma$ the relation (\ref{4.3}) is real. Namely, for $\sigma=0$ we understand
Eq. (\ref{4.3}) in the sense of the limit procedure, thus, in this case
$\Psi=\phi$; for $\sigma<0$ Eq. (\ref{4.3}) can be rewritten as
$\Psi=\frac{\tan{\left((-\sigma)^{\frac{1}{2}}\phi\right)}}{(-\sigma)^{\frac{1}{2}}}$.
In this section we shall construct solutions for the $d=n=1$ and
$d=2,\,\,n=0$ string theories as extensions of an arbitrary solution
$(\phi,\,\,h_{\mu\nu})$ of this effective dilaton gravity system.
Our extensions will preserve the asymptotical triviality of the seed
solutions in the following sense: dilaton gravity solutions with Coulomb
behavior at spatial infinity are mapped to string theory solutions
possessing the same underlying property. This means that we must perform
our extension procedure in such a way that all possible Dirac string
peculiarities, which naturally arise in the framework of any more or less
general symmetry based solution-constructing technique, will vanish.
For example, in General Relativity the Ehlers symmetry
transformation \cite{Ehl} generates the parameter NUT from the
asymptotically trivial Schwarzshild solution; the resulting family of Taub-NUT solutions \cite{EMT} possesses
a Dirac string peculiarity of rotational type. In the pure General
Relativity
it is impossible to specify the generation procedure in the above mentioned sense
because, in fact, there exists only a one-parametric charging symmetry
transformation
whose `specification' leads to an identical map. This transformation coincides with
the Ehlers symmetry which must be `normalized' to preserve the asymptotical flatness
property, see, for example, \cite{HKCS}, where this material together with its
straightforward generalization to the string theory case is considered in details.
Note that in our approach all the classical nonlinear symmetries (like the
Ehlers and Harrison transformations) arise in a matrix-valued framework;
nevertheless it is possible to identify some string theory symmetries as the Harrison
and Ehlers maps in the conventional non-matrix sense, see \cite{ind1}, \cite{ind2}.
It is clear that in the string theories
under consideration the group of charging symmetries has a
much more rich structure \cite{ZF}; moreover, it is possibile to relate
different string theories labeled by distinct values of the parameters $d$
and $n$ in a charging symmetry invariant form \cite{HSTI}. All these
facts
allows one to realize the above mentioned extension program not
only in the case of the two special string theories under consideration, but also for the
general case. Below we give the corresponding results for the $d=n=1$ and
$d=2,\,\,n=0$ theories.

The concrete asymptotically trivial and charged dilaton gravity solution
of Nordstrom -Reissner type reads
\be\label{4.4}
&&\phi=\frac{1}{\sqrt 2\sigma^{\frac{1}{2}}}\log\left
(\frac{R+m\sigma^{\frac{1}{2}}}
{R-m\sigma^{\frac{1}{2}}}\right ),
\nonumber\\
&&ds_3^2=dR^2+(R^2-\sigma m^2)(d\theta^2+\sin^2\theta d\varphi^2),
\ee
where $m$ is an arbitrary real parameter. We note again that the
$\phi$-relation
in Eq. (\ref{4.4}) is real and well defined for the arbitrary sign of $\sigma$: if
$\sigma=0$ then $\phi=\frac{\sqrt 2m}{R}$; for $\sigma<0$ one also has from Eq.
(\ref{4.4}) that $\phi=\frac{1}{\sqrt 2(-\sigma)^{\frac{1}{2}}}
{\rm arctan}\left (\frac{2m(-\sigma)^{\frac{1}{2}}R}{R^2+m^2\sigma}\right )$.
Thus, Eqs. (\ref{4.1}), (\ref{4.3})
and (\ref{4.4}) give a correct solution of Eqs. (\ref{3.11}) which
have, as it is easy
to see, the standard monopole behavior at spatial infinity. From Eqs. (\ref{4.3})
and (\ref{4.4}) it follows that the parameter $\sqrt 2m$ plays the role of the formal
Coulomb
charge of this solution. The solution (\ref{4.4}) gives a concrete realization of the
on--shell field configurations of the general charged dilaton gravity with
the trivial asymptotics at spatial infinity. In principle, our further
analysis can be related to such general asymptotically flat solutions.
However, we shall take into account just the special class of solutions
corresponding to the Nordstrom-Reissner one.

Our next step is to calculate the matrix potentials $\M_{\alpha}$ and
$\vec\Omega_{\alpha}$ according to Eq. (\ref{2.4}). In order to get the more symmetric form of the
resulting formulas let us redefine the parameter $\P$ as $\P\rightarrow\P\Sigma$; this
map preserves the $\tau$-value, see Eq. (\ref{4.2'}). The result reads:
\be\label{4.5}
&&\M_1=\Sigma+\frac{\sinh^2\left (\sigma^{\frac{1}{2}}\phi\right )}{\tau}\P^T\P,
\quad \vec\Omega_1=0;
\nonumber\\
&&\M_2=\frac{\sinh\left (\sigma^{\frac{1}{2}}\phi\right )
\cosh\left (\sigma^{\frac{1}{2}}\phi\right )}{\sigma^{\frac{1}{2}}}\P^T\Q,
\quad \vec\Omega_2=\vec\omega\P^T\Q;
\nonumber\\
&&\M_3=\frac{\sinh^2\left (\sigma^{\frac{1}{2}}\phi\right )}{\kappa}\Q^T\Q,
\quad \vec\Omega_3=0,
\ee
where the vector function $\vec\omega$ is defined on--shell by the
relation
\be\label{4.6}
\nabla\times\vec\omega=\nabla\phi.
\ee
The explicit calculation of $\vec\omega$ in the case of the solution (\ref{4.4}) gives
the following only non-zero component of this vector function
\be\label{4.6'}
\omega_{\varphi}=\sqrt 2m\cos\theta.
\ee
Of course, one obtains the same asymptotical value of $\vec\omega$
for the general solution of the dilaton gravity with a leading term
of monopole type.

Now let us compute the potentials $S_0$, $S_{\alpha}$ and $\vec
V_{\alpha}$ for the two
concrete string theories under consideration. For the $d=n=1$ theory $\P=(\P_1,\P_2)$ and
$\Q=(\Q_1,\Q_2,\Q_3)$ are $1\times 2$ and $1\times 3$ rows respectively, whereas
$\Sigma=-1$ is a $2\times 2$ matrix and $G_0=-1$ is a number. In this special case the
block segmentations given by Eq. (\ref{2.5}) coincide with the usual matrix structures
and all the calculations are especially simple. So, for the potentials $\vec V_{\alpha}$
one immediately obtains that
\be\label{4.7}
\vec V_1=\left (\P_1\Q_2-\P_2\Q_1\right )\vec\omega,\quad
\vec V_2=\left (\P_1\Q_2+\P_2\Q_1\right )\vec\omega,\quad
\vec V_3=\sqrt 2\P_1\Q_3\vec\omega.
\ee
These vector quantities lead to the appearance of Dirac string peculiarities for the
metric (the NUT parameter), gauge (the magnetic charge) and Kalb Ramond fields according
to Eqs. (\ref{2.7}) and (\ref{4.7}). In order to remove these
peculiarities and to obtain the
guaranteed asymptotically flat string theory solutions which correspond to the charged
solutions of the effective dilaton gravity system, one must restrict the parameters $\P$
and $\Q$ in such a way that all the quantities $\vec V_{\alpha}$ vanish identically. Let
us also keep the arbitraryness of the $\sigma$ value for the
restricted class of
solutions in order to preserve the possibility of working within
dilaton gravity with
arbitrary coupling. These conditions lead to a unique choice of the
restriction:
$\P_1=\Q_1=0$ and arbitrary values for the remaining parameters $\P_2,\Q_2$ and
$\Q_3$. In this special case the calculation of the scalar potentials becomes much
simpler and one finally obtains $S_0=1$ and
\be\label{4.8}
&&S_1=-\left [\cosh\left (\sigma^{\frac{1}{2}}\phi\right )+\Q_2
\frac{\sinh\left (\sigma^{\frac{1}{2}}\phi\right )}{(-\kappa)^{\frac{1}{2}}}\right ]^2,
\nonumber\\
&&S_2=-\left [\Q_3
\frac{\sinh\left (\sigma^{\frac{1}{2}}\phi\right )}{(-\kappa)^{\frac{1}{2}}}\right ]^2,
\nonumber\\
&&S_3=-\sqrt 2\Q_3
\frac{\sinh\left (\sigma^{\frac{1}{2}}\phi\right )}{(-\kappa)^{\frac{1}{2}}}
\left [\cosh\left (\sigma^{\frac{1}{2}}\phi\right )+\Q_2
\frac{\sinh\left (\sigma^{\frac{1}{2}}\phi\right )}{(-\kappa)^{\frac{1}{2}}}\right ],
\ee
where now $\tau=-\P_2^2$ and $\kappa=-\Q_2^2+\Q_3^2$. Note that the parameter $\P_2$
can be removed from Eqs. (\ref{4.4}) and (\ref{4.8}) with the help of the substitutions
$\sigma\rightarrow -\kappa$,\, $\phi\rightarrow |\P_2|\phi$ and $m\rightarrow |\P_2|m$.
Equations (\ref{4.8}) define the
extension of the arbitrary solution of the dilaton gravity system (\ref{4.2}) to the case
of the $d=n=1$ theory completely. This extension is automatically free of any string
peculiarity of Dirac type $(\vec V_{\alpha}= 0)$
(up to construction of this extension). In particular, the solution (\ref{4.4}) leads
to charged asymptotically trivial heterotic string theory fields.

Now let us consider the bosonic string theory case with $d=2,\,n=0$.
Here the
block structure of the matrix potentials $\M_{\alpha}$ and $\vec\Omega_{\alpha}$ is
not trivial, and it is convenient to parameterize the $1\times 3$ rows $\P$ and $\Q$
as $\P=(\P_1,\,P)$ and $\Q=(\Q_1,\,Q)$ where $P=(\P_2,\,\P_3)$ and
$Q=(\Q_2,\,\Q_3)$,
respectively. The calculation of the $\vec V_{\alpha}$-potentials gives the
following result:
\be\label{4.9}
\vec V_1=\left (\P_1Q^T-\Q_1\sigma_3P^T\right )\vec\omega,\quad
\vec V_2=\left (\P_1\sigma_3Q^T+\Q_1P^T\right )\vec\omega,
\ee
where $\sigma_3=-G_0$ in this theory. A further removal of the Dirac string
peculiarities from
the solution leads again to the restriction $\P_1=\Q_1=0$ and to arbitrary values of
$P$ and $Q$ under the same assumptions that in the previous case. Then, one obtains again
that $S_0=1$, whereas for the remaining part of the scalar potentials one has the following
relations:
\be\label{4.10}
&&S_1\!=\!-\!\sigma_3\!+\!\sinh^2\left (\sigma^{\frac{1}{2}}\phi\right )
\left [\frac{P^TP}{\tau}\!+\!\frac{Q^TQ}{\kappa}\right ]\!-\!
\frac{\sinh\left (\sigma^{\frac{1}{2}}\phi\right )
\cosh\left (\sigma^{\frac{1}{2}}\phi\right )}{\sigma^{\frac{1}{2}}}
\left [ P^TQ\!+\!QP^T\right ],\\
&&S_2\!=\!
\sinh^2\left (\sigma^{\frac{1}{2}}\phi\right )
\left [\!-\frac{P^TP}{\tau}\!+\!\frac{Q^TQ}{\kappa}\right ]\sigma_3\!+\!
\frac{\sinh\left (\sigma^{\frac{1}{2}}\phi\right )
\cosh\left (\sigma^{\frac{1}{2}}\phi\right )}{\sigma^{\frac{1}{2}}}
\left [ Q^TP\!-\!PQ^T\right ]\sigma_3,\nonumber
\ee
where now $\tau=-P\sigma_3P^T$ and
$\kappa=-Q\sigma_3Q^T$ (we have also performed the map
$P\rightarrow\sigma_3P$ which
preserves the $\tau$-value and leads to some simplifications of the
result). The last step
is to calculate the quantities ${\rm det}\, S_1$ and $S_1^{-1}$. For ${\rm det}\, S_1$, after
some algebra, one concludes that
\be\label{4.11}
{\rm det}\, S_1=-\left [\cosh\left (\sigma^{\frac{1}
{2}}\phi\right )+\frac{\sinh\left (\sigma^{\frac{1}{2}}\phi\right )}{\sigma^
{\frac{1}{2}}}Q\sigma_3P^T\right ]^2,
\ee
whereas for $S_1^{-1}$ one simply obtains
$S_1^{-1}=\frac{1}{{\rm det}\, S_1}\sigma_2S_1^T\sigma_2$ in view of its $2\times 2$
matrix nature. Eqs. (\ref{4.10}), (\ref{4.11}) together with the relations
$\vec V_{\alpha}= 0$ and the
above calculated quantity $S_1^{-1}$ completely define the $d=2,\,n=0$
bosonic string theory extension of the effective three-dimensional dilaton gravity.
Again, up to construction, this extension is automatically free of any Dirac string
peculiarity. This last step completes the program formulated at the end of the previous
section;
any further analysis will be related to the special choice of the solution of the
dilaton gravity system of equations (\ref{4.2}). In this context we refer to the
Nordstrom-Reissner type solution given by Eq. (\ref{4.4}) as to the typical monopole solution
which is widely represented in the classical $\sigma$-model gravity theories.

\section{Conclusion}
\setcounter{equation}{0}

The results of this article allows one to transform static electromagnetic
solutions of the Einstein-Maxwell theory to the corresponding solutions of the
five-dimensional bosonic string theory and also to extend the electric (magnetic)
static Einstein-Maxwell fields to the four-dimensional heterotic string theory with one
vector field. As an alternative starting theory for the generating
procedure it is possible to use the stationary
General Relativity. Let us note that in our approach all the classical starting
systems arise in some interpretation invariant form. This means that we do not actually
start from, for example, General Relativity which represents the string theory ansatz
with vanishing matter fields and Kaluza-Klein metric modes. In fact our Einstein-Maxwell
and General Relativity subsystems arise
as some
formal objects from the symmetry invariant point of view and their physical nature is
nothing else than a very special choice of the starting theories in the
framework of our
formalism.
The same situation takes
place in the effective three-dimensional dilaton gravity, which we have
explored in
details as the natural and the simplest physically interesting starting system. In
particular, we have presented explicit relations for the extension of the solutions of
the dilaton gravity system with arbitrary value of coupling to both the heterotic and
bosonic string theories under consideration. More precisely, we have
shown that this
extension can be performed in a form which is free of any peculiarity
of Dirac string type for solutions with a leading term of
monopole type at
spatial infinity. Note that the
removal of these peculiarities means in fact some gauge fixing
with respect to the subgroup of
charging symmetries.  This subgroup forms a total invariance
class of extensions which are free of any
parameter fixing. Actually, the general dilaton gravity extension scheme
is related to
arbitrary parameters $\P$ and $\Q$, which effectively transform as
$\P\rightarrow\P\Sigma C_1$ and $\Q\rightarrow\Q C_2$ under the
charging symmetry subgroup of transformations
(\ref{2.8}). However, the consistent removal of peculiarities demands the
vanishing of the first components of the rows $\P$ and $\Q$ - a condition which
does not hold when applying general charging symmetry transformations.
Thus, the charging symmetry invariant class of string theory charged solutions related to the
dilaton gravity subsystem definitely contains Dirac string peculiarities, whereas the
completely asymptotically trivial string theory field configuration is charging
symmetry non-invariant.

As it was briefly mentioned in the Introduction and also partially
supported by our Nordstrom-Reissner
solution example, the natural and the nearest
applications of the new solution generation procedure proposed in this
article is
black hole physics. We hope to generalize our approach to the case of
string theory with arbitrary dimensionality and with arbitrary number of the Abelian
gauge fields. The leading principle for this generalization will be related to demanding
a continuous and charging symmetry invariant extension of the general
Israel-Wilson-Perjes class of solutions \cite{ZF} to the non-extremal case. In a
forthcoming publication \cite{m2} we hope to realize this program and to reach,
in particular, the
critical cases for the heterotic and bosonic string theories, which are the most
interesting from the physical point of view.


\section*{Acknowledgments}
The work of both authors was supported by CONACyT grant $N^0$\,\,$J34245-E$;
the work of A.H.-A. was also supported by grant CIC-$4.18$ whereas the work of
O.V.K. - by grant RFBR $N^0$\,\,$00\,02\,17135$. O.V.K. thanks IFM UMICH for
facilities and hospitality provided during his stay at Morelia, Michoacan,
Mexico.


\end{document}